\documentclass[conference]{IEEEtran}
\IEEEoverridecommandlockouts

\usepackage{cite}
\usepackage{amsmath,amssymb,amsfonts}
\usepackage{algorithmic}
\usepackage{graphicx}
\usepackage{textcomp}
\usepackage{xcolor}
\usepackage{booktabs}
\usepackage{makecell}
\usepackage{float}
\usepackage{utfsym}
\usepackage{pifont}
\usepackage{graphicx} 
\usepackage{caption}
\usepackage{mathptmx}
\usepackage{url}
\usepackage{enumitem}

\definecolor{mygrey}{cmyk}{0, 0, 0, 1}
\newcommand{\smallgrey}[1]{\textcolor{mygrey}{\scalebox{0.9}{#1}}}
\def\BibTeX{{\rm B\kern-.05em{\sc i\kern-.025em b}\kern-.08em
    T\kern-.1667em\lower.7ex\hbox{E}\kern-.125emX}}
\begin{document}

\title{SFNet: A Spatial-Frequency Domain Deep Learning Network for Efficient Alzheimer's Disease Diagnosis\\
}

\author{
\IEEEauthorblockN{1\textsuperscript{st} Xinyue Yang}
\IEEEauthorblockA{\textit{School of Artificial Intelligence}\\ \textit{Beijing Normal University } \\
Beijing, China\\
202321081046@mail.bnu.edu.cn}
\and
\IEEEauthorblockN{2\textsuperscript{nd} Meiliang Liu}
\IEEEauthorblockA{\textit{School of Artificial Intelligence} \\
\textit{Beijing Normal University}\\
Beijing, China \\
liumeiliang520@gmail.com}
\and
\IEEEauthorblockN{3\textsuperscript{rd} Yunfang Xu}
\IEEEauthorblockA{\textit{School of Artificial Intelligence} \\
\textit{Beijing Normal University}\\
Beijing, China \\
xumouse222@gmail.com}
\and
\IEEEauthorblockN{4\textsuperscript{th} Xiaoxiao Yang}
\IEEEauthorblockA{\textit{School of Artificial Intelligence} \\
\textit{Beijing Normal University}\\
Beijing, China \\
791690301@qq.com}
\and
\IEEEauthorblockN{5\textsuperscript{th} Zhengye Si}
\IEEEauthorblockA{\textit{School of Artificial Intelligence} \\
\textit{Beijing Normal University}\\
Beijing, China \\
sizhengye0302@gmail.com}
\and
\IEEEauthorblockN{6\textsuperscript{th} Zijin Li}
\IEEEauthorblockA{\textit{School of Artificial Intelligence} \\
\textit{Beijing Normal University}\\
Beijing, China \\
2291832685@qq.com}
\and
\IEEEauthorblockN{7\textsuperscript{th} Zhiwen Zhao}
\IEEEauthorblockA{\textit{School of Artificial Intelligence} \\
\textit{Beijing Normal University}\\
Beijing, China \\
mlt.bnu2017@bnu.edu.cn}
}
\maketitle

\begin{abstract}
Alzheimer's disease (AD) is a progressive neurodegenerative disorder that predominantly affects the elderly population and currently has no cure. Magnetic Resonance Imaging (MRI), as a non-invasive imaging technique, is essential for the early diagnosis of AD. MRI inherently contains both spatial and frequency information, as raw signals are acquired in the frequency domain and reconstructed into spatial images via the Fourier transform. However, most existing AD diagnostic models extract features from a single domain, limiting their capacity to fully capture the complex neuroimaging characteristics of the disease. While some studies have combined spatial and frequency information, they are mostly confined to 2D MRI, leaving the potential of dual-domain analysis in 3D MRI unexplored. To overcome this limitation, we propose Spatial-Frequency Network (SFNet), the first end-to-end deep learning framework that simultaneously leverages spatial and frequency domain information to enhance 3D MRI-based AD diagnosis. SFNet integrates an enhanced dense convolutional network to extract local spatial features and a global frequency module to capture global frequency-domain representations. Additionally, a novel multi-scale attention module is proposed to further refine spatial feature extraction. Experiments on the Alzheimer's Disease Neuroimaging Initiative (ADNI) dataset demonstrate that SFNet outperforms existing baselines and reduces computational overhead in classifying cognitively normal (CN) and AD, achieving an accuracy of 95.1\%.
\end{abstract}

\section{INTRODUCTION}
Alzheimer’s disease (AD) is an irreversible neurodegenerative disorder characterized by the progressive deterioration of memory and cognitive functions \cite{b1}. Due to the lack of effective pharmacological treatments, early diagnosis and intervention are essential for patients \cite{b2}. Magnetic resonance imaging (MRI), as a non-invasive neuroimaging technique with high spatial resolution, is valuable for identifying structural brain alterations associated with AD \cite{b3}. However, manual analysis of MRI in clinical practice is often time-consuming and subject to inter-observer variability. Consequently, deep learning-based methods have been increasingly explored for the automatic detection of AD from MRI data. These approaches offer the potential to enhance diagnostic accuracy and efficiency, thereby facilitating more reliable clinical decision-making.

Among various deep learning architectures, convolutional neural networks (CNNs) have shown considerable effectiveness in AD classification tasks\cite{b4}. For instance, Korolev et al. \cite{b5} proposed a 3D neural network that combined VGGNet and ResNet for AD diagnosis. Fan et al. \cite{b6} applied the U-Net to AD classification, enhancing the diagnostic accuracy. Wang et al.\cite{b7} proposed an ensemble of 3D densely connected convolutional networks with a probability-based fusion method to boost the performance of AD diagnosis.

To further extract spatial features from MRI, spatial attention and channel attention have been widely proposed \cite{b8,b9,b10,b11,b12,b37,b38}. Gao et al.\cite{b8} proposed a multi-scale attention convolution to learn feature maps with multi-scale kernels. Dutta et al. \cite{b9} combined spatial attention and self-attention blocks in parallel to comprehensively capture the feature dependencies along spatial dimensions. Liu et al.\cite{b10} proposed a multi-plane and multi-scale feature-level fusion attention model, which used a multi-scale feature extractor with hybrid attention layers to simultaneously capture and fuse multiple pathological features in the sagittal, coronal, and axial planes. Zhang et al. \cite{b11} integrated two squeeze-and-excitation (SE) blocks to embed channel attention into a multi-scale fusion (MSF) feature extraction module. Tang et al.\cite{b12} introduced a spatial channel attention module (ECSA), which can focus on important AD-related features in images more efficiently.

In addition, the self-attention mechanism introduced by the Transformer \cite{b13} has exhibited remarkable performance across various computer vision applications \cite{b14,b15}. However, the direct application of the Vision Transformer (ViT) to MRI entails two significant challenges. First, the self-attention mechanism is computationally intensive, with a complexity of $O(n^2)$ \cite{b16}. Second, MRI datasets are generally much smaller than those commonly used in computer vision, limiting the effective training of ViT. Therefore, numerous studies have explored the integration of Transformer with CNN to improve AD classification while reducing computational costs \cite{b17,b18,b19,b20,b21}. Li et al. \cite{b17} introduced Trans-ResNet, which combined ResNet-18 for spatial feature extraction with a Transformer to capture long-range dependencies. Jang et al. \cite{b18} proposed a hybrid model that integrated 3D CNN, 2D CNN, and Transformer for AD diagnosis, achieving an accuracy of 93.21\%. Hu et al. \cite{b19} proposed Conv-Swinformer, which utilized VGG-16 for low-level spatial feature extraction and employed a sliding window strategy to fuse adjacent features. Khatri et al. \cite{b20} incorporated convolution-attention mechanisms within Transformer-based classifiers to enhance performance while maintaining computational efficiency. Miao et al.\cite{b21} proposed a Multi-modal Multi-scale Transformer Fusion Network (MMTFN) for AD diagnosis. The multi-scale features were extracted from each modality using 3D multi-scale residual blocks and fused via the Transformer.

While the above models reduced computational cost to some extent by incorporating CNN and Transformer, they did not fundamentally address the $O(n^2)$ complexity brought by the self-attention mechanism. To tackle this issue, Rao et al. \cite{b22} proposed the Global Filter Network (GFNet), which substituted the self-attention with the fast Fourier transform. This approaches achieved a computational complexity of $O(n \log n)$ while effectively capturing long-term spatial dependencies in the frequency domain. Building on this idea, Zhang et al. \cite{b23} extended the model to a 3D Global Fourier Network and applied it to extract long-range dependencies in MRI of AD patients. Furthermore, Kushol et al. \cite{b24} proposed ADDformer, which leveraged transfer learning to train ViT and GFNet with a majority voting strategy for classification, resulting in an accuracy of 88.2\%.

Despite significant progress achieved by frequency domain models, several limitations remain. First, certain models, such as GFNet, exclusively extracted features from the frequency domain while neglecting the spatial structural information inherent in MRI data. Second, although some models integrated spatial and frequency domain features, they often rely on 2D slices, resulting in a lack of spatial continuity across the three-dimensional volume. To overcome these limitations, we propose a 3D deep learning model that integrates spatial and frequency domain features to fully exploit the complementary information embedded in MRI data. In the spatial domain, an improved DenseNet architecture is employed to extract local structural features, enhanced by a multi-scale spatial attention module to improve the perception of brain regions at different spatial resolutions and scales. In the frequency domain, a global frequency module applies the fast Fourier transform to feature maps, capturing long-range dependencies across regions and enabling effective global context modeling. By fusing spatial-frequency domain features, the model enhances inter-regional interactions and significantly improves its discriminative performance in predicting early-stage Alzheimer's disease. The main contributions of this paper are as follows:

\begin{itemize}
    \item We propose a novel model for AD diagnosis, termed SFNet, which is the first to integrate spatial local features and global frequency-domain dependencies based on 3D MRI data, effectively enhancing the accuracy of AD classification.
    \item A multi-scale attention module is proposed in the spatial domain to expand the receptive field and capture multi-scale local spatial features.
    \item We employ a low-rank MLP layer in the frequency domain, allowing the model to reduce model parameters and computation. Furthermore, learnable global filters are visualized to improve model interpretability by revealing spectral responses.
\end{itemize}
\section{MATERIALS}
\subsection{Datasets}
The raw MRI images utilized in this study are obtained from the Alzheimer's Disease Neuroimaging Initiative (ADNI, http://adni.loni.usc.edu). We employ T1-weighted structural MRI scans (MPRAGE) from a total of 1484 participants. Specifically, in the ADNI-1 and ADNI-2 datasets, we collect 1.5 Tesla MRI scans from 1222 participants, including 415 cognitively normal (CN) individuals, 477 with mild cognitive impairment (MCI), and 330 diagnosed with AD. For the ADNI-3 dataset, we obtain 3 Tesla MRI scans from 262 participants, comprising 150 AD patients and 112 CN individuals. The demographic characteristics of each cohort are summarized in Table~\ref{information_label}.

\begin{table}[htbp]
    \centering
    \caption{Demographic information of the subjects in the study. Age and MMSE scores are presented as mean$\pm$standard deviation (SD).}
    \renewcommand{\arraystretch}{1.4} 
    \resizebox{0.479\textwidth}{!}{ 
    \begin{tabular}{cccc}
    \toprule
    \textbf{Type} & \makecell{\textbf{Gender}\\\textbf{(Male/Female)}} & \makecell{\textbf{Age}\\\textbf{(Mean $\pm$ SD)}} & \makecell{\textbf{MMSE}\\\textbf{(Mean $\pm$ SD)}} \\
    \midrule

    CN  & 233/294 & 72.66\smallgrey{$\pm$6.39} & 29.07\smallgrey{$\pm$1.15} \\
    MCI & 193/284 & 73.50\smallgrey{$\pm$7.36} & 27.86\smallgrey{$\pm$1.87} \\
    AD  & 216/264 & 76.06\smallgrey{$\pm$7.87} & 22.78\smallgrey{$\pm$2.86}  \\
    \bottomrule
    \end{tabular}
    }
    \label{information_label}
\end{table}
\subsection{Image Preprocessing}
\begin{figure}[h]
    \centering
    \includegraphics[width=0.47\textwidth]{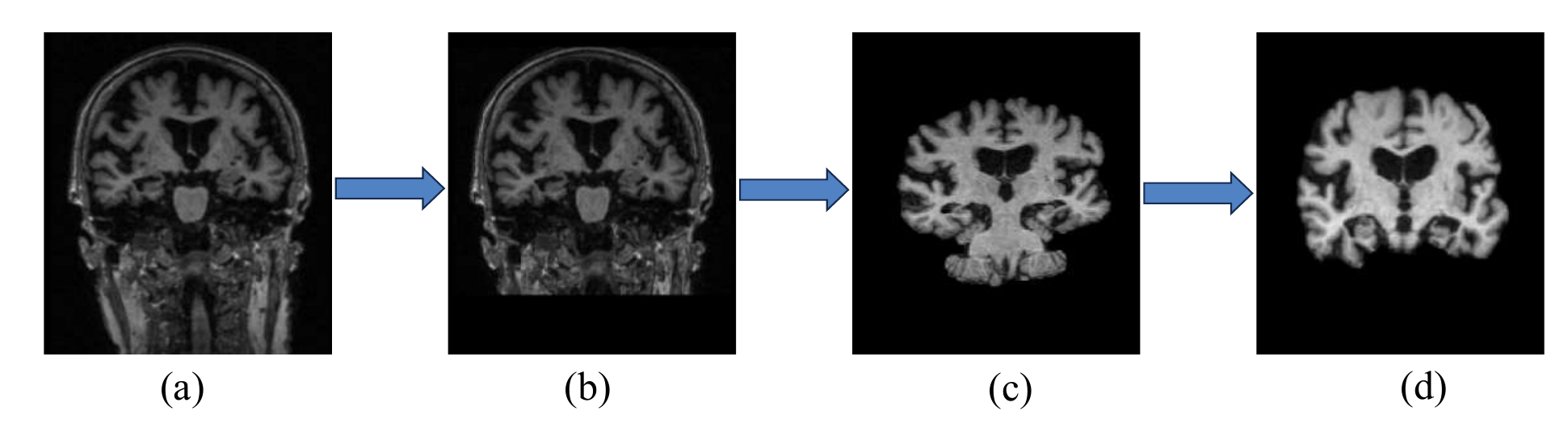}
    \caption{The MRI scans preprocessing pipeline (coronal plane). (a) Original MRI scans; (b) Neck removal; (c) Skull stripping; (d) Linear registration.}
    \label{preprocess_fig}
\end{figure}

All MRI scans are preprocessed using a standardized pipeline, as illustrated in Figure~\ref{preprocess_fig}. First, DICOM files are converted to NIfTI (Neuroimaging Informatics Technology Initiative) format using the SimpleITK library. Then, neck removal is conducted using the FSL (FMRIB's Software Library) toolbox~\cite{b25} to eliminate non-cranial structures. Skull stripping is conducted using the Brain Extraction Tool (BET)~\cite{b26}, resulting in brain-only images. Finally, each brain image is linearly registered to the MNI152-1mm space using FLIRT (FMRIB’S linear image registration tool). Following these steps, all MRI scans are resampled to a uniform voxel resolution of $182\times218\times182$.

\begin{figure*}[h]
    \centering
    \includegraphics[width=0.85\textwidth]{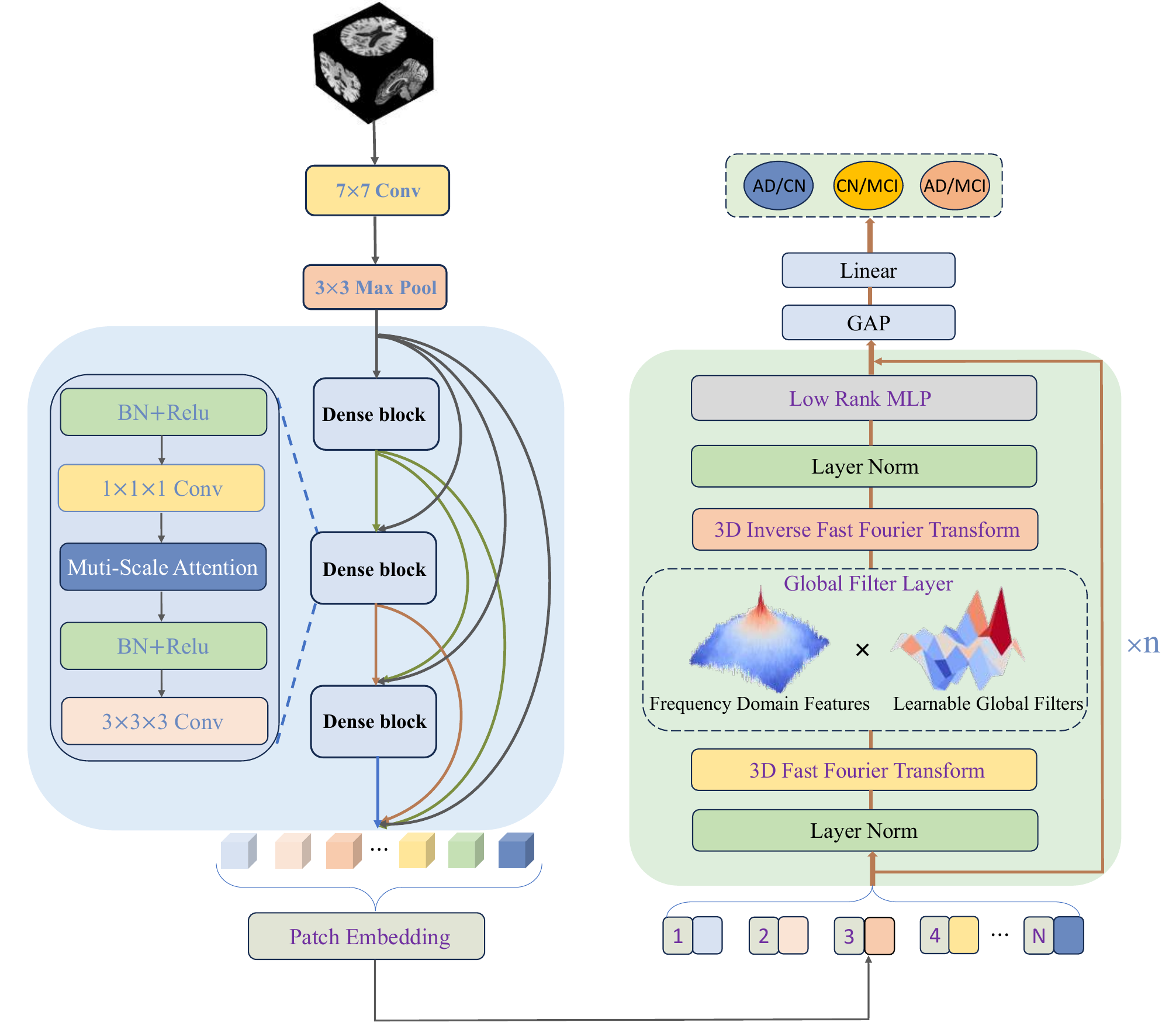}
    \caption{The architecture of SFNet}
    \label{network_fig}
\end{figure*}

\section{Network Architecture}
The architecture of the proposed SFNet is depicted in Figure~\ref{network_fig}. Initially, the input MRI scans pass through a $7\times7\times7$ convolutional layer for preliminary feature extraction, followed by a $3\times3\times3$ max pooling layer to reduce spatial dimensions. The extracted features are subsequently fed into the spatial feature extraction module to capture multi-scale local patterns, and then into the global frequency module to model long-range dependencies across brain regions. Finally, a fully connected layer classifies subjects into AD/CN, AD/MCI, or CN/MCI. Detailed descriptions of each module are as follows:

\subsection{Spatial Feature Extraction Module}
The spatial feature extraction module consists of two components. First, DenseNet is employed as the backbone network to extract spatial features. Then, a novel multi-scale attention module is incorporated to progressively expand the receptive field and capture multi-scale local patterns.
\subsubsection{Spatial convolution}
The dense block was first proposed in the DenseNet~\cite{b27} by adding the dense direct connections between standard convolution layers in a feed-forward fashion. For a dense block comprising $L$ layers, the output of the $l^{\text{th}}$ layer, $x_l$, is defined as:
\begin{equation}
 \label{equ_1}
 x_l = H_l[x_0, x_1, \ldots, x_{l-1}]
\end{equation}
where $H_l(\cdot)$ is denoted a nonlinear transformation, and $[x_0, x_1, \ldots, x_{l-1}]$ represents the concatenation of the feature maps generated by layers $0$ through $l-1$. These dense connectivities facilitate efficient feature reuse and mitigate the vanishing-gradient problem.

Inspired by DenseNet, we adapt the principle of dense connectivity for AD classification. To enhance spatial feature extraction, an attention module is introduced between the two convolutional layers within each dense block, while retaining the dense connections between blocks. Specifically, each dense block begins with a $1\times1\times1$ bottleneck convolution for dimensionality reduction, followed by the attention mechanism. Subsequently, a $3\times3\times3$ convolution is employed to produce 3D feature representations. Batch Normalization (BN) and ReLU activation are applied prior to each convolutional layer. The transformation within a dense block can be expressed as:
\begin{equation}
    x_l=Conv_{3\times 3\times 3} (\phi(A(Conv_{1\times1\times1}(\phi(x_{l-1})))))
\end{equation}
where $\phi(\cdot)$ denotes BN followed by a ReLU activation, and $A(\cdot)$ represents the attention module.

\subsubsection{Multi-scale Attention Module}
\begin{figure}[h]
    \centering
    \includegraphics[width=1\columnwidth]{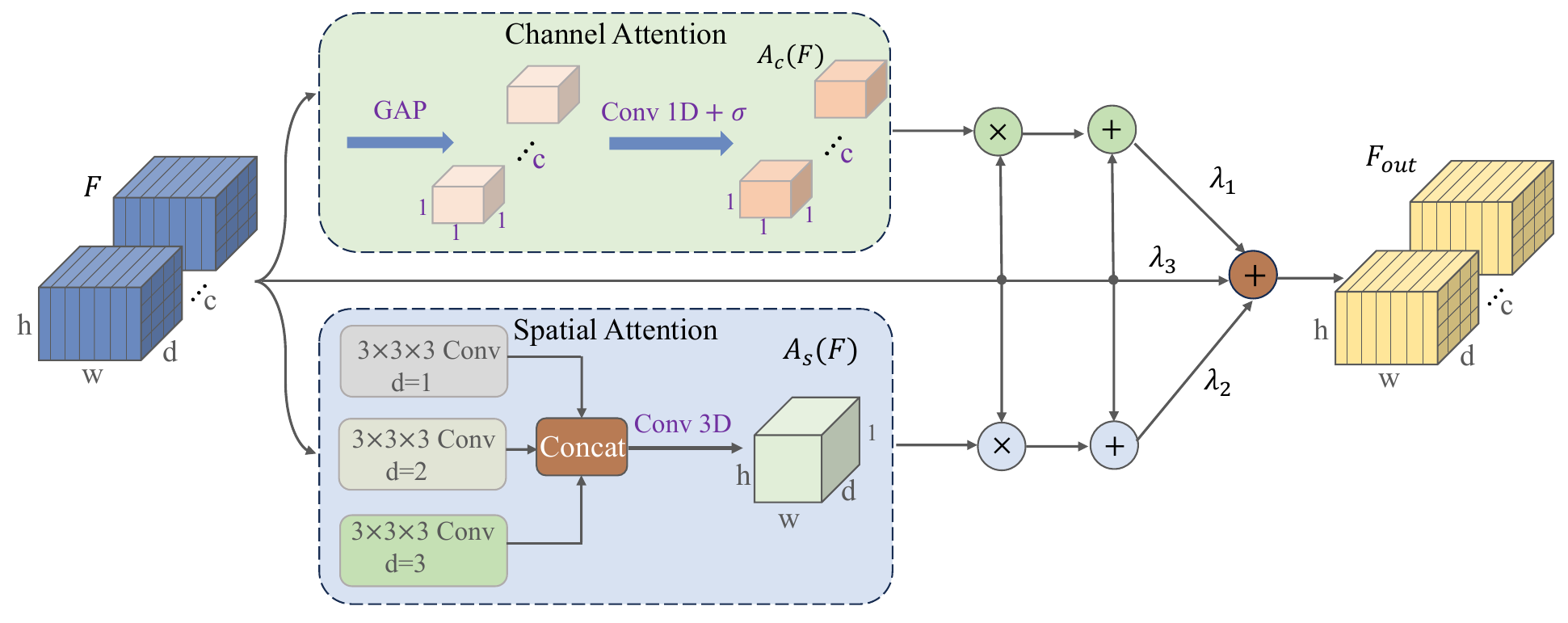}
    \caption{Multi-scale attention module}
    \label{attention_fig}
\end{figure}

To further enhance feature extraction from the spatial domain, we propose a multi-scale attention module,as shown in Figure~\ref{attention_fig}.

The module adopts a parallel architecture comprising two branches: a channel attention branch and a spatial attention branch. The former enhances inter-channel feature discrimination, while the latter employs dilated convolutions to capture local spatial information at multiple scales. To preserve original features and promote efficient gradient flow, residual connections are embedded in both branches. By jointly modeling complementary channel and spatial representations, this module plays a crucial role in strengthening spatial feature extraction.
\paragraph{Channel Attention}
The channel attention aims at appropriately capturing local cross-channel interaction by considering every channel and its $k$ neighbors. Let the input feature map be $F\in \mathbb{R}^{C\times W\times H\times D}$, where $C, W, H, D$ represent the number of channels, width, height, and depth of the feature map, respectively. The attention map in the channel domain, $A_c(F)$, is denoted as:
\begin{equation}
    A_c(F)=\sigma\left(Conv_{1D}(AvgPool(F))\right) \in \mathbb{R}^{C\times 1\times 1\times 1}
    \label{equ_2}
\end{equation}
where $AvgPool$ is the global average pooling (GAP), $Conv_{1D}$ is $1D$ convolution of kernel size $k$, and $\sigma$ is the Sigmoid activation function. To adapt the convolutional kernel size to the channel dimension $C$, we adopt a nonlinear mapping $C = 2^{\gamma \cdot k - b}(k=2,b=1)$, which allows $k$ to vary with $C$ in a logarithmic fashion, following the design of ECA-Net~\cite{b28}.
\paragraph{Spatial Attention}

Dilated convolution facilitates multiscale feature extraction by incorporating varying dilation rates within the same convolutional kernel, thereby expanding the receptive field without increasing the number of parameters. This property is particularly advantageous in AD diagnosis, where it is essential to capture both fine-grained and large-scale spatial patterns. In the spatial domain, we apply dilated convolution to the input $F\in \mathbb{R}^{C\times W\times H\times D}$, producing feature maps $F_i$ as follows:
\begin{equation}
F_i = Conv_{d=i}(Relu(BN(F))), \quad i=1,2,3
\end{equation}
where $Conv_{d=i}$ represents the $3\times 3\times 3$ dilated convolution with dilation rates of 1, 2, and 3 respectively.

Then, the feature maps $F_i$ are concatenated and sent to a $3D$ convolution layer to generate the spatial attention map $A_s(F)$, which is denoted as: 
\begin{equation}
A_s(F) = Conv_{3d}(Concat(F_1, F_2, F_3))\in \mathbb{R}^{1\times W\times H\times D}
\end{equation}

Finally, the output of the attention module is determined by the attention maps produced by the two branches. 
\begin{equation}
    F_{out} = \lambda_1 (F \odot A_c(F) + F) + \lambda_2 (F \odot A_s(F) + F) + \lambda_3 F
\end{equation}
where \( \odot \) denotes element-wise multiplication, \( + \) inside the parentheses represents element-wise addition, and \( \lambda_1, \lambda_2, \lambda_3 \) are learnable weights initialized to 0.33. Furthermore, residual connections enhance the feature extraction capability of the attention module by integrating the original features with the attention-weighted representations.

\subsection{Global Frequency Module}
The combination of spatial convolution and multi-scale attention enables the model to detect subtle changes associated with disease effectively. However, these operations primarily focus on local patterns and multi-scale features, potentially neglecting the global connections between distant brain regions. To address this limitation, we introduce a global frequency module, including fast Fourier transforms, global filter layers, and inverse fast Fourier transforms, as shown in Figure~\ref{network_fig}.

Prior to frequency domain operations, the output of the spatial feature extraction module, $X \in \mathbb{R}^{\tilde{C} \times \tilde{W} \times \tilde{H} \times \tilde{D}}$, is divided into non-overlapping 3D patches of size $P \times P \times P$. Each patch is then flattened and linearly projected into a vector of dimension $D = \tilde{C} \times P \times P \times P$, resulting in a total of $L = \frac{\tilde{W}}{P} \times \frac{\tilde{H}}{P} \times \frac{\tilde{D}}{P}$ tokens. For each image, these tokens are stacked to form $\tilde{X} \in \mathbb{R}^{L \times D}$, which serves as the input to the global frequency module.


\begin{table*}[h] 
\centering
\caption{Model comparison on ADNI dataset}
\renewcommand{\arraystretch}{1.8} 
\begin{tabular}{lccccccccccccc}
\toprule
\textbf{Method} 
& \multicolumn{4}{c}{\textbf{AD vs CN }} 
& \multicolumn{4}{c}{\textbf{CN vs MCI }} 
& \multicolumn{4}{c}{\textbf{AD vs MCI}}
& \textbf{Params(M)}\\
\cmidrule(lr){2-5} \cmidrule(lr){6-9} \cmidrule(lr){10-13}
& ACC & SEN & SPE & AUC 
& ACC & SEN & SPE & AUC 
& ACC & SEN & SPE & AUC \\
\midrule
ResNet-18    & 0.921 & 0.906 & 0.934 & 0.981 
             & 0.714 & 0.727 & 0.693 & 0.721       
             & 0.751 & 0.689 & 0.846 & 0.761 & 33.16\\
3D GFNet     & 0.846 & 0.885 & 0.817 & 0.923
             & 0.785 & 0.890 & 0.659 & 0.860 
             & 0.756 & 0.838 & 0.654 & 0.801 & 70.40\\
ViT          & 0.915 & 0.894 & 0.939 & 0.918
             & 0.708 & 0.695 & 0.748 & 0.712 
             & 0.714 & 0.629 & 0.812 & 0.748 & 27.31\\
M3T          & 0.929 & 0.930 & \textbf{0.955} & 0.934
             & 0.755 & 0.741 & \textbf{0.820} & 0.766
             & 0.785 & 0.795 & \textbf{0.885} & 0.779 &29.12 \\
Conv-Swinformer & 0.935 & 0.938 & 0.933 & 0.975
                & 0.791 & 0.798 & 0.782 & 0.858
                & 0.821 & 0.869 & 0.714 & 0.857 & 24.30\\
CViT         & 0.921 & 0.924 & 0.951 & 0.930
             & 0.742 & 0.710 & 0.785 & 0.746
             & 0.782 & 0.782 & 0.873 & 0.766 & 15.90\\
SFNet     & \textbf{0.951} & \textbf{0.959} & 0.939 & \textbf{0.984}
             & \textbf{0.860} & \textbf{0.924} & {0.784} & \textbf{0.917}
             & \textbf{0.849} & \textbf{0.884} & 0.807 & \textbf{0.917} &18.75 \\
\bottomrule
\end{tabular}
\label{all_result}
\end{table*}

We first transform $\tilde{X}$ into the spectral domain using a 3D fast Fourier transform. Then, the global filter, a key component for capturing long-range interactions in the frequency domain, performs element-wise multiplication with learnable filters to generate $\hat{X}$.
\begin{equation}
    \hat{X}=K\odot\mathcal{F}[\tilde{X}]
\end{equation}
where $\mathcal{F}$ is the 3D fast Fourier transform, $\odot$ is the element-wise multiplication, $K$ is the learnable global filter which has the same dimension as $X$. Finally, the 3D inverse fast Fourier transform restores the modulated spectrum $\hat{X}$ to the spatial domain, yielding updated tokens $\tilde{X'}\in \mathbb{R}^{D\times L}$.
\begin{equation}
    \tilde{X'}\xleftarrow{}\mathcal{F}^{-1}[\tilde{X}]   
\end{equation}

Following the inverse fast Fourier transform, we apply a two-layer low-rank MLP to further transform the features. Different from the standard MLP layer, which employs full-rank linear projections, the linear layers in our MLP are decomposed using a low-rank approximation to reduce parameters and computational costs, which is denoted as:
\begin{equation}
    MLP(\tilde{X'})=\phi(W_4W_3\cdot(\phi(W_2W_1(\tilde{X'}))))
\end{equation}
where \( W_1\in \mathbb{R}^{D \times r_1} \), \( W_2 \in \mathbb{R}^{r_1 \times D'} \), \( W_3 \in \mathbb{R}^{D' \times r_2} \), \( W_4 \in \mathbb{R}^{r_2 \times D} \), and \( \phi(\cdot) \) denotes GELU activation. The output of the MLP layer is fed into a classification head for prediction, which consists of a GAP layer followed by a FC layer.

\section{Experiments and Discussions}

\subsection{Experimental Settings}
All experiments are implemented in Python 3.8 and PyTorch 2.4.0, and conducted on an NVIDIA A40 GPU with 48 GB of memory. A five-fold cross-validation strategy is adopted to assess the model's performance across three classification tasks: AD/CN, AD/MCI, and CN/MCI. Specifically, 15\% of the entire dataset is reserved as the test set, while the remaining 85\% is used for training and validation. Model training is performed using the AdamW optimizer with an initial learning rate of 0.0005, scheduled via cosine annealing. The model is trained for 130 epochs with a batch size of 18, employing the cross-entropy loss function. The dense block stacks 3 times, and the depth of the global frequency module is set to 6.

The performance of the model is evaluated using four metrics: accuracy (ACC), sensitivity (SEN), specificity (SPE), and F1-score. These evaluation metrics are defined as follows:
\begin{equation}
    Accuracy=\frac{TP+TN}{TP+TN+FP+FN}
\end{equation}
\begin{equation}
    Sensitivity=\frac{TP}{TP+FN}
\end{equation}
\begin{equation}
    Specificity=\frac{TN}{TN+FP}
\end{equation}
\begin{equation}
    F1-score=\frac{2\times TP}{2\times TP+FN+FP}
\end{equation}
where TP, FN, FP, and TN are denoted as true positive, false negative, false positive, and true negative, respectively. Additionally, we employ the area under the receiver operating characteristic curve (AUC) to assess the model’s overall discriminative ability. A higher AUC reflecting a better trade-off between sensitivity and specificity across a range of classification thresholds.

\subsection{Model Comparsion On All Three Classification Tasks}
\label{result_analysis}
In this section, we compare the proposed SFNet with recent related methods for the classification of AD/CN, CN/MCI, and AD/MCI, as summarized in Table~\ref{all_result}. The results show that SFNet outperforms GFNet, highlighting the effectiveness of incorporating spatial-frequency domain information for feature extraction. Compared to Transformer-based models, such as ViT, as well as CNN-Transformer-based models like M3T and CViT, SFNet consistently demonstrates superior performance. Specifically, it achieves higher ACC, SEN, and AUC while maintaining a lower number of parameters. This superiority stems from SFNet’s ability to capture long-range dependencies in the frequency domain, which more effectively complements spatial features than the spatial self-attention mechanisms used in CNN-Transformer-based models.
Notably, compared to the best-performing Conv-Swinformer method, SFNet reduces model complexity and achieves improvements of 1.6\% in AD/CN classification accuracy, 6.9\% in CN/MCI, and 2.8\% in AD/MCI.

Furthermore, as MCI represents an intermediate stage between CN and AD, classification tasks involving MCI are inherently more challenging. Nevertheless, our model achieves notable improvements in both CN/MCI and AD/MCI classifications. We attribute this to two key factors. First, different from models restricted to a single domain, our approach jointly leverages local spatial features and long-range frequency dependencies, thereby enhancing representational capacity. Second, the spatial feature maps incorporate multi-scale spatial attention, channel attention, and the original image, allowing the model to capture both high-level semantic and fine-grained structural information. This design enhances feature discriminability under a low-parameter regime, which is particularly advantageous for identifying subtle differences in intermediate stages such as MCI.

\subsection{Comparison With Advanced Methods For AD/CN Classification}
In section \ref{result_analysis}, SFNet is compared with several state-of-the-art models for AD/CN, CN/MCI, and AD/MCI classification. Furthermore, to provide a comprehensive evaluation, we extend the comparison to another nine models, including AddFormer~\cite{b24}, DA-MIDL~\cite{b29}, sMRI-PatchNet~\cite{b30}, MRN~\cite{b31}, CNN+Aging Transformer~\cite{b32},  LA-GMF~\cite{b33}, LongFormer~\cite{b34}, LiCoL~\cite{b35}, and CE-AH~\cite{b36}. These methods employ various input strategies, including slice-based, region-based, and 3D MRI-based techniques. To ensure a fair comparison, their performances are directly extracted from the respective original publications. The comparative results are presented in Table~\ref{table_other}.

\begin{table}[h]
\centering
\caption{Comparison with recent advanced deep learning methods for AD/CN classification}
\renewcommand{\arraystretch}{1.8} 
\resizebox{1\columnwidth}{!}{
\begin{tabular}{llcccc}
\toprule
\textbf{Reference} & \textbf{Method} & \textbf{ACC} & \textbf{SEN} & \textbf{SPE} & \textbf{AUC} \\
\midrule
DA-MIDL (2021)               & patch-based             & 0.924 & 0.910 & 0.938 & 0.965\\
sMRI-PatchNet (2023)         & patch-based             & 0.920 & 0.920 & 0.919 & 0.967\\
CNN+Aging Transformer (2023) & 3D MRI-based            & 0.905 & 0.846 & 0.950 & 0.939\\
AddFormer (2022)             & slice-based             & 0.882 & 0.956 & 0.774 & - \\
MRN (2023)                   & regions-based           & 0.925 & 0.833 & \textbf{0.966} & 0.976\\
LA-GMF (2024)                & 3D MRI-based            & 0.930 & 0.939 & 0.923 & 0.949 \\
LongFormer (2024)            & 3D MRI-based            & 0.934 &-      &-      &0.933\\
CE-AH (2025)                 & Multi-scale level based & 0.949 &0.955  &0.943  &0.973\\
LiCoL (2025)                 & 3D MRI-based            & 0.928 & 0.926 & 0.958 & 0.936 \\
SFNet                        & 3D MRI-based            &\textbf{0.951}  & \textbf{0.959} &0.939 &\textbf{0.984}\\
\bottomrule
\end{tabular}
}
\label{table_other}
\end{table}


Specifically, AddFormer utilizes slice-based input, whereas DA-MIDL and sMRI-PatchNet employ patch-based strategies. Although these approaches effectively capture local structural information, they might ignore essential global contextual features. MRN achieves competitive specificity but suffers from reduced sensitivity and accuracy due to its reliance on anatomical priors. Although models such as CNN+Aging Transformer, LA-GMF, LongFormer, LiCoL, and CE-AH also leverage the entire 3D MRI as input, their performance still lags behind that of the proposed SFNet. Specifically, SFNet achieves the highest accuracy (0.951), sensitivity (0.959), and AUC (0.84), highlighting its superior capability in capturing comprehensive spatial-frequency representations from 3D MRI data.

\subsection{Ablation Experiments}
The ablation studies are conducted to evaluate the individual contributions of each component within SFNet to overall performance, deriving insights into the impact of the model's fundamental elements on classification. Firstly, we conduct an ablation experiment for three classification tasks, investigating the impacts of spatial convolution, multi-scale attention, and global frequency module on SFNet performance. Specifically, we designed three variants: baseline, baseline + spatial convolution, and baseline + spatial convolution + multi-scale attention. The baseline model employs the global frequency module.

\begin{figure}[h]
    \centering
    \includegraphics[width=0.5\textwidth]{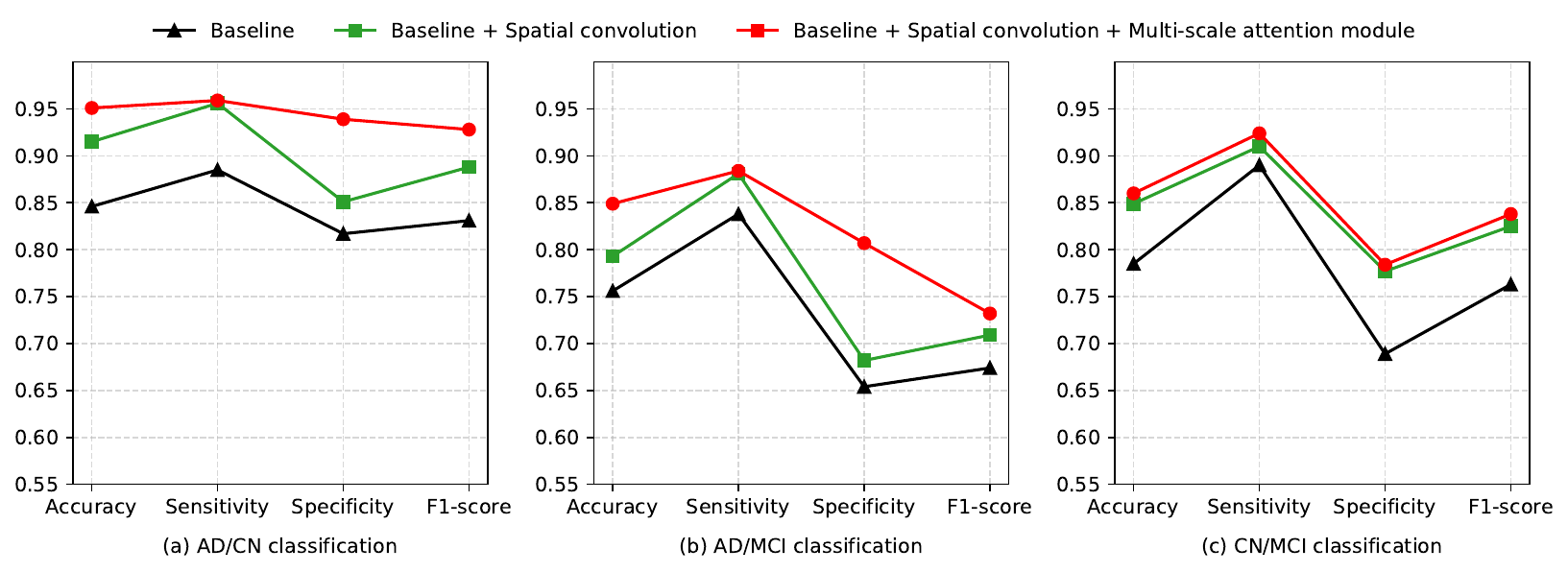}
    \caption{Ablation experiments results on the three classification tasks}
    \label{ablation_all}
\end{figure}

As illustrated in Figure~\ref{ablation_all}, the baseline model achieves relatively balanced performance in the AD/CN classification task (ACC=0.846$\pm$0.02, SEN=0.885$\pm0.04$, SPE=0.817$\pm0.02$, F1-score=0.831$\pm0.02$). However, its performance declined notably in AD/MCI and CN/MCI tasks, particularly in terms of SPE, which drops to 0.654$\pm$0.04 and 0.659$\pm$0.05, respectively.
The incorporation of spatial convolution significantly improves the performance of all three classification tasks, particularly in the AD/CN, where it achieves an ACC of 0.915$\pm$0.03, SEN of 0.956$\pm$0.02, SPE of 0.851$\pm$0.04, and an F1-score of 0.888$\pm$0.03. This indicates that combining both the spatial and frequency domains is especially effective in enhancing overall classification accuracy and sensitivity. Furthermore, the integration of the multi-scale attention yields additional improvements among the three classification tasks. Importantly, it also significantly enhances the specificity of AD classification tasks, increasing it from 0.682$\pm$0.05 to 0.80$\pm$0.02 in AD/MCI classification tasks, demonstrating a strong ability to identify AD cases. Overall, these results demonstrate the effectiveness of spatial convolution and multi-scale attention in enhancing classification accuracy, as well as their respective contributions to improving sensitivity and specificity.

\begin{figure}[h]
    \centering
    \includegraphics[width=0.477\textwidth]{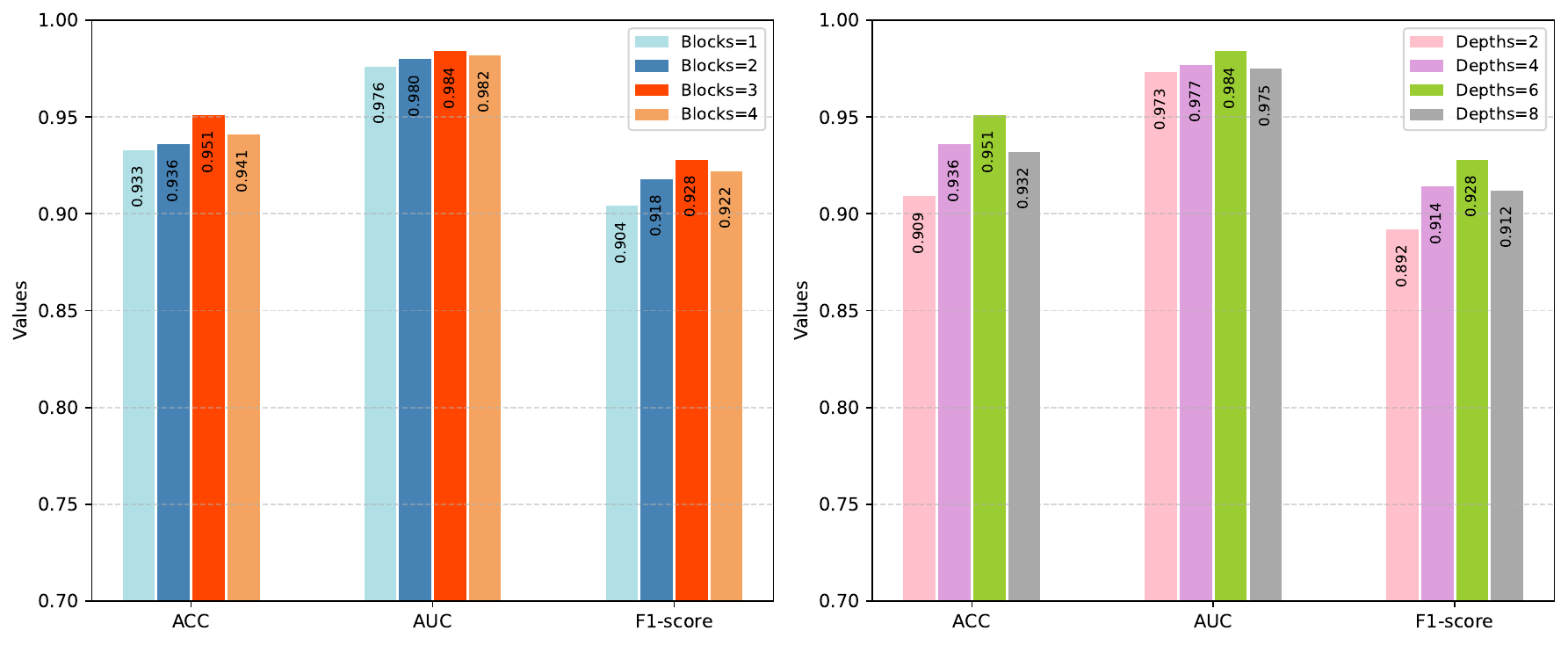}
    \caption{Performance of SFNet with different dense blocks and global frequency depths}
    \label{Ablation_hyperparameters}
\end{figure}

The second ablation study examines how the number of dense blocks and the depth of the global frequency module influence the performance and computational complexity of SFNet. As illustrated in Figure~\ref{Ablation_hyperparameters} and Table~\ref{ablation_params}, increasing both the number of dense blocks and the depth of the frequency module enhances classification accuracy and F1-score. However, the number of dense blocks contributes more significantly to computational cost compared to the frequency module depths. Based on these findings, a configuration comprising three dense blocks and a global frequency module depth of six is selected to achieve an optimal trade-off between performance and efficiency.

\begin{table}[h]
\centering
\caption{Params and FLOPs affected by dense blocks number and global frequency module depths}
\renewcommand{\arraystretch}{1.4} 
\resizebox{1\columnwidth}{!}{
\begin{tabular}{cccc}
\toprule
\makecell{\bf Dense blocks \\ \bf number} & \makecell{\bf Global frequency \\\bf module depths} &\textbf{Params (M)} &\textbf{FLOPs (G)} \\
\midrule
Blocks=1 & Depths=6 &5.80  &11.91\\
Blocks=2 & Depths=6 &11.3  &19.87\\
Blocks=3 & Depths=6 &18.75 &29.48\\
Blocks=4 & Depths=6 &27.92 &40.75\\
Blocks=3 & Depths=2 &9.01 &20.79\\
Blocks=3 & Depths=4 &13.88 &25.14\\
Blocks=3 & Depths=8 &23.62 &33.82\\
\bottomrule
\end{tabular}}
\label{ablation_params}
\end{table}

\begin{figure*}[ht!]
    \centering
    \includegraphics[width=1\textwidth]{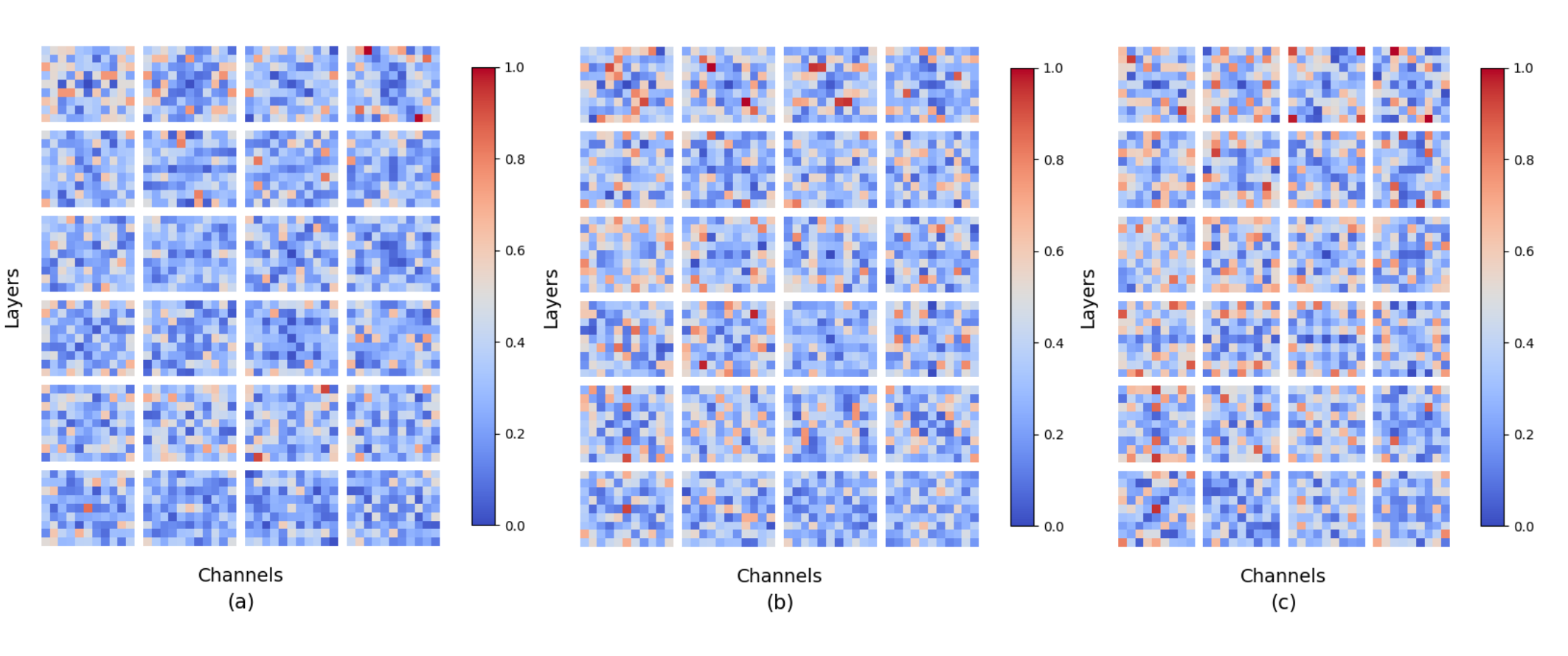}
    \caption{Learnable global filters visualization. (a): Direction XY; (b): Direction YZ; (c): Direction XZ}
    \label{filter_xy}
\end{figure*}

\subsection{Frequency Domain Visualization}
As learnable weights, the learnable global filters in the global frequency module reveal how the model attends to different frequency components, thereby providing interpretability of the SFNet. As shown in Figure~\ref{filter_xy}, we visualize four representative channels selected from the filter along three orthogonal planes (a: direction xy, b: direction yz, c: direction xz) based on the AD/CN classification task. Each image represents the frequency response intensity of the filter weights after performing the Fourier transform and centralization, where the center of each patch corresponds to low-frequency components and the edges represent high-frequency components. The visualizations are generated by slicing the 3D filter weights along three axes and arranging them by layer and channel.

In the frequency domain, the learned global filters exhibit distinct patterns across different network layers. Consistently across all three anatomical views, earlier layers demonstrate stronger activations in high-frequency regions. In contrast, deeper layers progressively shift toward emphasizing low-frequency components or exhibit a more uniform frequency response. For instance, in the first channel, the final layer shows its strongest activation in the central region of the spectrum, indicating a clear focus on low-frequency information. This progressive transition from high- to low-frequency emphasis reflects the network's hierarchical feature extraction process, wherein earlier layers capture fine-grained details and deeper layers abstract more global representations, consistent with the general behaviour of deep neural networks. Furthermore, noticeable differences among the three views suggest that the model effectively captures direction-specific frequency characteristics based on the anatomical slicing plane. This observed directional and hierarchical modulation enhances the interpretability of the global frequency module.

\section{Conclusions}
In this paper, we propose an end-to-end deep learning framework, SFNet, for the early diagnosis of Alzheimer’s disease using 3D MRI data. The model incorporates a multiscale attention module and DenseNet for spatial feature extraction, alongside the global frequency module to capture frequency-domain information. This dual-domain approach enables SFNet to effectively learn both local fine-grained textures and global structural patterns. Experiments on ADNI datasets demonstrate that SFNet outperforms recent state-of-the-art methods in classification accuracy while maintaining a lower total parameters. Furthermore, we visualize the learnable global filters in the frequency domain that reveal how SFNet progressively adapts its attention to different spectral components, contributing to the interpretability of the model. In future work, we plan to focus on multi-model integration to enhance the robustness and generalizability of SFNet.


\begin{thebibliography}{00}
\bibitem{b1} Better, MAPPING A. "Alzheimer’s disease facts and figures." Alzheimers Dement 19.4 (2023): 1598-1695.
\bibitem{b2} Huang L K, Chao S P, Hu C J. Clinical trials of new drugs for Alzheimer disease[J]. Journal of biomedical science, 2020, 27: 1-13.
\bibitem{b3} Frisoni G B, Fox N C, Jack Jr C R, et al. The clinical use of structural MRI in Alzheimer disease[J]. Nature reviews neurology, 2010, 6(2): 67-77.
\bibitem{b4} Wen J, Thibeau-Sutre E, Diaz-Melo M, et al. Convolutional neural networks for classification of Alzheimer's disease: Overview and reproducible evaluation[J]. Medical image analysis, 2020, 63: 101694.
\bibitem{b5} Korolev S, Safiullin A, Belyaev M, et al. Residual and plain convolutional neural networks for 3D brain MRI classification[C]//2017 IEEE 14th international symposium on biomedical imaging (ISBI 2017). IEEE, 2017: 835-838.
\bibitem{b6} Fan Z, Li J, Zhang L, et al. U-net based analysis of MRI for Alzheimer’s disease diagnosis[J]. Neural Computing and Applications, 2021, 33: 13587-13599.
\bibitem{b7} Wang H, Shen Y, Wang S, et al. Ensemble of 3D densely connected convolutional network for diagnosis of mild cognitive impairment and Alzheimer’s disease[J]. Neurocomputing, 2019, 333: 145-156.
\bibitem{b8} Gao X, Cai H, Liu M. A hybrid multi-scale attention convolution and aging transformer network for Alzheimer's disease diagnosis[J]. IEEE Journal of Biomedical and Health Informatics, 2023, 27(7): 3292-3301.
\bibitem{b9} Dutta T K, Nayak D R, Zhang Y D. Arm-net: Attention-guided residual multiscale cnn for multiclass brain tumor classification using mr images[J]. Biomedical Signal Processing and Control, 2024, 87: 105421.
\bibitem{b10} Liu F, Wang H, Liang S N, et al. MPS-FFA: A multiplane and multiscale feature fusion attention network for Alzheimer’s disease prediction with structural MRI[J]. Computers in Biology and Medicine, 2023, 157: 106790.
\bibitem{b11} Zhang L, Xia R, Yang B, et al. MSFNet‐2SE: A multi‐scale fusion convolutional network for Alzheimer's disease classification on magnetic resonance images[J]. International Journal of Imaging Systems and Technology, 2024, 34(4): e23112.
\bibitem{b12} Tang C, Xi M, Sun J, et al. MACFNet: Detection of Alzheimer's disease via multiscale attention and cross-enhancement fusion network[J]. Computer Methods and Programs in Biomedicine, 2024, 254: 108259.
\bibitem{b37} Liu M, Xu Y, Li Z, et al. Kolmogorov-Arnold Networks for Time Series Granger Causality Inference[J]. arXiv preprint arXiv:2501.08958, 2025.
\bibitem{b38} Liu M, Dong H, Yang X, et al. Gradient Regularization-based Neural Granger Causality[J]. arXiv preprint arXiv:2507.11178, 2025.
\bibitem{b13} Vaswani A, Shazeer N, Parmar N, et al. Attention is all you need[J]. Advances in neural information processing systems, 2017, 30.
\bibitem{b14} Khan, Salman, et al. "Transformers in vision: A survey." ACM computing surveys (CSUR) 54.10s (2022): 1-41.
\bibitem{b15} Carion, Nicolas, et al. "End-to-end object detection with transformers." European conference on computer vision. Cham: Springer International Publishing, 2020.
\bibitem{b16} Zhu J, Tan Y, Lin R, et al. Efficient self-attention mechanism and structural distilling model for Alzheimer’s disease diagnosis[J]. Computers in Biology and Medicine, 2022, 147: 105737.
\bibitem{b17} Li C, Cui Y, Luo N, et al. Trans-resnet: Integrating transformers and cnns for alzheimer’s disease classification[C]//2022 IEEE 19th International Symposium on Biomedical Imaging (ISBI). IEEE, 2022: 1-5.
\bibitem{b18} Jang J, Hwang D. M3T: three-dimensional Medical image classifier using Multi-plane and Multi-slice Transformer[C]//Proceedings of the IEEE/CVF conference on computer vision and pattern recognition. 2022: 20718-20729.
\bibitem{b19} Hu Z, Li Y, Wang Z, et al. Conv-Swinformer: Integration of CNN and shift window attention for Alzheimer’s disease classification[J]. Computers in Biology and Medicine, 2023, 164: 107304.
\bibitem{b20} Khatri U, Kwon G R. Diagnosis of Alzheimer's disease via optimized lightweight convolution-attention and structural MRI[J]. Computers in Biology and Medicine, 2024, 171: 108116.
\bibitem{b21} Miao S, Xu Q, Li W, et al. MMTFN: Multi‐modal multi‐scale transformer fusion network for Alzheimer's disease diagnosis[J]. International Journal of Imaging Systems and Technology, 2024, 34(1): e22970.
\bibitem{b22} Rao Y, Zhao W, Zhu Z, et al. Global filter networks for image classification[J]. Advances in neural information processing systems, 2021, 34: 980-993.
\bibitem{b23} Zhang S, Chen X, Ren B, et al. 3d global fourier network for alzheimer’s disease diagnosis using structural mri[C]//International Conference on Medical Image Computing and Computer-Assisted Intervention. Cham: Springer Nature Switzerland, 2022: 34-43.
\bibitem{b24} Kushol R, Masoumzadeh A, Huo D, et al. Addformer: Alzheimer’s disease detection from structural mri using fusion transformer[C]//2022 IEEE 19th International Symposium on Biomedical Imaging (ISBI). IEEE, 2022: 1-5.
\bibitem{b25} Jenkinson M, Beckmann C F, Behrens T E J, et al. Fsl[J]. Neuroimage, 2012, 62(2): 782-790.
\bibitem{b26} Isensee F, Schell M, Pflueger I, et al. Automated brain extraction of multisequence MRI using artificial neural networks[J]. Human brain mapping, 2019, 40(17): 4952-4964.
\bibitem{b27} Huang G, Liu Z, Van Der Maaten L, et al. Densely connected convolutional networks[C]//Proceedings of the IEEE conference on computer vision and pattern recognition. 2017: 4700-4708.
\bibitem{b28} Wang Q, Wu B, Zhu P, et al. ECA-Net: Efficient channel attention for deep convolutional neural networks[C]//Proceedings of the IEEE/CVF conference on computer vision and pattern recognition. 2020: 11534-11542.
\bibitem{b29} Zhu W, Sun L, Huang J, et al. Dual attention multi-instance deep learning for Alzheimer’s disease diagnosis with structural MRI[J]. IEEE Transactions on Medical Imaging, 2021, 40(9): 2354-2366.
\bibitem{b30} Zhang X, Han L, Han L, et al. sMRI-PatchNet: A novel efficient explainable patch-based deep learning network for Alzheimer’s disease diagnosis with Structural MRI[J]. IEEE Access, 2023, 11: 108603-108616.
\bibitem{b31} Zhang J, He X, Qing L, et al. Multi-relation graph convolutional network for Alzheimer’s disease diagnosis using structural MRI[J]. Knowledge-Based Systems, 2023, 270: 110546.
\bibitem{b32} Gao X, Cai H, Liu M. A hybrid multi-scale attention convolution and aging transformer network for Alzheimer's disease diagnosis[J]. IEEE Journal of Biomedical and Health Informatics, 2023, 27(7): 3292-3301.
\bibitem{b33} Xu J, Yuan C, Ma X, et al. Interpretable medical deep framework by logits-constraint attention guiding graph-based multi-scale fusion for Alzheimer’s disease analysis[J]. Pattern Recognition, 2024, 152: 110450.
\bibitem{b34} Chen Q, Fu Q, Bai H, et al. Longformer: longitudinal transformer for Alzheimer's disease classification with structural MRIs[C]//Proceedings of the IEEE/CVF winter conference on applications of computer vision. 2024: 3575-3584.
\bibitem{b35} Oh K, Heo D W, Mulyadi A W, et al. A quantitatively interpretable model for Alzheimer’s disease prediction using deep counterfactuals[J]. NeuroImage, 2025: 121077.
\bibitem{b36} Wang T, Dai Q, Lu H. CE-AH: A Contrast-Enhanced Attention Hierarchical Network for Alzheimer's Disease Diagnosis Based on Structural MRI[J]. Pattern Recognition, 2025: 111986.
\end{thebibliography}
\end{document}